# Dispersive Casimir Pressure Effect from Surface Plasmon Quanta by Quasi 1D Metal Wires in Ferrite Disks and The Josephson Frequencies and Currents

## Mahmut Obol

Ferrites are distinct material for electromagnetic applications due to its unique spin precession. In this paper, Casimir pressure effect by deploying magnetically tunable surface plasmon quanta in stratified structure of using ferrite and metal wires is presented. Previously, oscillating surface plasmon quanta were successfully included to modify first reflection and first transmission characteristics. The oscillating surface plasmon quanta in the modified reflection in such a system, not only does resolve in a typical matter in metamaterial, but also provide new applications such as creating Casimir pressure effects through the metamaterial composite shown in this paper. The Casimir pressure flips from attractive state to repulsive state is referred to cause mechanism of radiation from surface plasmon quanta. Both Casimir force analysis and the measured data of radiations indicate us the system develops quantized states by electric flux induced by ferromagnetic resonance. Quantum analysis is used to understand the discrete radiations spectra for our experimental measurement. The discrete radiations are reproduced by using time dependent Schrödinger representation. As result, we find the Josephson frequency and Josephson current representations at room temperature and we used them for extrapolating voltage induced in excited ferrites. Josephson frequency at X-band is able to differentiate micron volt differences and it allows us to report the data for voltage induced by ferromagnetic resonance in ferrite at room temperature. It is understood that the radiation intensity depends on density of final states and excitation probability when we come to think the energy matter. It seems possible to create as high as 20mW microwave power inside waveguide at X-band.

I.  INTRODUCTION

The concept of negative refractive index material (NIM) was introduced in 1968 by Veselago [1] and a fabricated metamaterial composite was reported in 2000 by Smith et al. [2]. The study of NIM has been very popular recently due to its association with extraordinary electromagnetic characteristics including superlenses [3]. In general, the electromagnetic properties of a normal material are defined by permeability $\mu = \mu_r - j\mu_i$ and permittivity $\varepsilon = \varepsilon_r – j\varepsilon_i$. In NIM, the permeability and permittivity [4], change their sign from positive to negative, that is, $-\mu = -\mu_r + j\mu_i$ and $-\varepsilon = -\varepsilon_r + j\varepsilon_i$. It appears that a normal material with a positive absorption corresponds to negative absorption in NIM. Although such unusual appearance in metamaterials has been understood by using causality [5], such negative absorption is beyond the concepts of classical electrodynamics [4].

Recently, the theoretical concept of a ferrite with metal wires involving metamaterial composite was reported by Rachford et al. [6]. We applied this concept by fabricating a similar metamaterial composite of metal wires and ferrites, and were able to demonstrate a characteristic measurement in terms of permeability and permittivity. We immediately encountered apparent negative absorption phenomena. We introduced surface plasmon quanta [7] in our metamaterial composite, which removes atypical phenomenon of negative absorption. The measured permeability and permittivity can be expressed differently from before as follows; $-\mu = -\mu_r - j\mu_i$ and $-\varepsilon = -\varepsilon_r - j\varepsilon_i$. These show both real negative permeability and real negative permittivity for NIM. The absorption components of metamaterial however, remain the same as in normal materials. It uses no casualty principles throughout our measurement process. This normalization comes with the inclusion of surface plasmon quanta [7] in NIM.

In this study, a metamaterial composite constructed by metal wires and ferrites disks was used. Here ferrites were responsible for creating negative permeability and the metal wires were responsible for inducing negative permittivity [8]. We shall refer to this metamaterial composite of metal wires and ferrite disks as MWIFD in this paper. The specific metal wire configuration alone eliminates the electromagnetic wave propagation through MWIFD because of the plasmonic metal wire structure inside the waveguide [8]. The nullified electromagnetic wave propagation does get recovered when spin precessions are driven around the metal wires by RF in the presence of an external magnetic field. Evidently, metal wires inside MWIFD composite, gain induced transient currents in the presence of RF and spin precessions in excited ferrites.

To find a solution or a derivation of Maxwell equations for the system thus becomes quite a complex, if one cannot distinguish these currents appropriately. Although numerous numerical Maxwell solvers are available, simultaneous dispersive permeability and permittivity characteristics present enormous challenges to using those simulation codes in negative refractive index media [4]. Despite those challenges, finite difference time domain simulation techniques were successfully implemented in a ferrite based negative index medium [6]. Most recently, a simultaneous permeability and permittivity measurement of ferrite was also reported by [9]. We noted that magnetically excited ferrites not only create dispersive permeability characteristics but also create dispersive permittivity within the ferromagnetic resonance spectrum. This is attributed to induced electric flux along the magnetization axis which creates ferromagnetic resonance in ferrites and shows an excellent agreement with recent work [10], where voltage enhancement is seen by uniform ferromagnetic resonance excitation in a yttrium iron garnet. Also, in a recent report [11], O. Mosendz et al. showed the voltage induced by inverse spin hall effect and the voltage induced were shown by using the terms of AC spin current and DC spin current. Authors [11] noted that the voltage induced at ferromagnetic resonance spectrum is proportional to the sample length. The electric flux induced by ferromagnetic resonance [9] is also proportional to the sample length along its magnetization axis and we noted that this finding also finds very good agreement with a recent report [11].

To properly understand internal wave mechanism of propagating mode in MWIFD, this paper introduces a dielectric vector potential concept into the system of MWIFD. In doing so, the induced electric flux contributions from spin flow precessions in excited ferrites in MWIFD can interact with metal wires perfectly by theoretical consideration. This extends magnetostatic limit into levels of Maxwell equations due to spins precession around the metal wires as well as the magnetization axis of the magnetic medium. This enables us to construct an additional set of Maxwell equations, which can recover the evanescent surface plasmon mode by using spin flow along RF wave propagation direction in the excited ferrite disks. The recreated wave propagation in MWIFD composite is defined as a cross product of two different evanescent modes in this study. Namely Feynman's notation [12] was used to derive propagation of such surface plasmons in MWIFD successfully. Then the propagating surface plasmon quanta from evanescent waves were used to derive Casimir pressure effects. Thus, it is natural for us to propose Casimir pressure effects in either dielectric or magnetic mirrors. Obtained Casimir pressure in this study it does flip its signs from an attractive state to a repulsive state when its refractive index and reflectivity do so. The obtained discrete Casimir forces and the obtained discrete radiation spectrum indicate us system develops quantized states by voltage induced by ferromagnetic resonance. As such we used the Schrödinger representation to derive Josephson frequency and Josephson current representations at room temperature. By using the Josephson frequencies, we are able to extrapolate the voltage induced by ferromagnetic resonance in YIG, and it implies that we are able to differentiate micron voltage differences in the system. When we come to think radiation intensity issues in terms of energy matter, we realized that the density of states is important. We present details of this study in sections below.

## II. THE **MWIFD** COMPOSITE IN WAVEGUIDE AND DIELECTRIC VECTOR POTENTIAL

Ferrite disks transversely magnetized ferrite disk to a perpendicular direction of wave propagation in a rectangular waveguide have been known for a long time. Normally, ferrites are used to understand Maxwell's equations and equation of motion for spin precessions. However, a composite of MWIFD inside a waveguide is difficult to describe with Maxwell's equations due to spin precession around the metal wires (see Fig.1). In this experiment, we apply an external static magnetic field along Y axis. The electromagnetic wave $TE_{10}$ propagates along Z-axis. Our experimental observation shows that in this configuration the inserted metal wires in ferrite disks are fully capable of eliminating the electromagnetic wave propagation in waveguide in the absence of an external magnetic field. An appropriate external magnetic field along the Y-axis is applied to the composite perpendicular to the wave propagation. This process recovers certain spectrum of propagation of nullified electromagnetic wave through the waveguide. This is a complex system, because it posses RF electromagnetic waves and spin precession associated waves with additional internal magnetic fields [13, 14, 15, and 16]. This study does not specify the spin precessions associated waves either from nonreciprocal surface waves or reciprocal bulk waves from excited ferrite; one may refer to the report [14] for detailed analysis of how to differentiate them.

We now present expression for effective permittivity of a surface plasmon state of this system is shown in equation (1) and the expression for plasmonic [8] resonance frequency as well as permittivity of nine parallel aligned metal wires in a unit cell (Fig.2) is shown in equation (2). The metal wires were considered as a negative permittivity medium using the MWIFD composite in our formulation process. A multilayered non homogeneous made up medium is loaded inside waveguide, see fig.1. In order to continuously use the $TE_{10}$ mode for this configuration inside waveguide, the configuration needs a medium homogenization constraint as shown below $\frac{k_{fx}}{n_f^2} = -\frac{k_{mx}}{n_m^2}$. One easily writes propagation constants for ferrite medium and plasmonic medium inside waveguide, they are $k_{mz}^2 + k_{mx}^2 = n_m^2\left(\frac{\omega}{c}\right)^2$ and $k_{fz}^2 + k_{zx}^2 = n_f^2\left(\frac{\omega}{c}\right)^2$. So, one obtains $k_{mz} = k_{fz} = k_{sp0} = \left(\frac{\omega}{c}\right)\sqrt{\frac{n_m^2 n_f^2}{n_m^2 + n_f^2}}$ and this configuration allows us to write the evanescent wave inside waveguide as follows.

$$\psi_{sp} = e^{\pm jk_{sp0}z}, \quad k_{sp0} = \frac{\omega}{c}\sqrt{\varepsilon_{sp}}, \text{ and } \varepsilon_{sp} = \frac{\varepsilon_m \varepsilon_f}{\varepsilon_f + \varepsilon_m \mu_f^{-1}} \quad (1)$$

where

$$\omega_m^2 = \frac{n_{meff} e^2}{\varepsilon_0 m_{eff}} = \frac{2\pi c_0^2}{a^2 \ln(b/d)} \text{ and } \varepsilon_m = 1 - \left(\frac{\omega_m}{\omega}\right)^2 \quad (2)$$

Moreover, ferrite disks in such system have spin flow precessions. Usually spin flow associated waves can be replaced by an internal magnetic field [13], $h_d$ and there is a requirement known as magnetostatic limit $\nabla \times \vec{h}_d = 0$. The magnetostatics associated waves must be seen if ferrites do not have metal wires in them. In a recent study for a YIG ferrite [9], we introduced an induced electric flux phenomenon due to the ferromagnetic resonance excitation in ferrite. Ferrites are insulator materials, so we do not expect electric charge and electric polarization ($\vec{p}_d$) build up to the ferrite surfaces due to the induced electric flux by ferromagnetic resonance in ferrites. So, the induced electric flux ($\vec{d}_m$) [9] satisfies the follows $\nabla \cdot \vec{d}_m = 0$ and $\nabla \cdot \nabla \times \vec{A}_d = 0$, where $\vec{d}_m = \nabla \times \vec{A}_d$ and $\vec{d}_m = \vec{e}_d + 4\pi \cdot (\vec{p}_d = 0) = \vec{e}_d$. By introducing a dielectric vector potential which implies that it also associates harmonic oscillation component in time, so it allows us to write an induced magnetic fields which referred to the contributions from uniform spin excitation and spin flow. As result, we have $h_d = -\frac{1}{c}\frac{\partial A_d}{\partial t}$, where $A_d$ stands for dielectric vector potential.

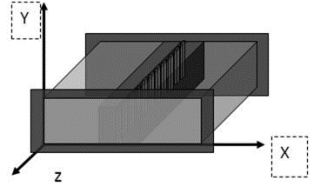

Fig. 1 A composite by metal wires and ferrite disks is located at the center position inside a rectangular waveguide.

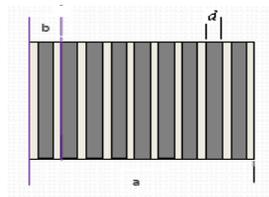

Fig. 2 Metal wires of aluminum foil placed in array formation inside the ferrite disk.

Specifically, dielectric vector potential itself alone may represent current loop without source, but it may be considered as sources of internal fields by spin flow and induced electric flux in excited ferrites. Correspondingly, we have an additional set of Maxwell equations for such expansion in this study, as shown below.

$$\nabla \times \vec{h}_d = -\frac{1}{c}\frac{\partial}{\partial t}\nabla \times \vec{A}_d = j\frac{\omega}{c}\nabla \times \vec{A}_d = j\frac{\omega}{c}\vec{e}_d \qquad (3)$$

$$\nabla \times \vec{e}_d = -j\frac{\omega}{c}[\tilde{\mu}] \cdot \vec{h}_d \qquad (4)$$

where $\tilde{\mu}$, Polder permeability tensor. By applying principles of rectangular waveguide, we also easily derive to flow of surface wave as follows.

$$\psi_\mu = e^{\pm jk_\mu z}, k_\mu = \frac{\omega}{c}\sqrt{\mu_f}, \text{ and } \mu_f = \frac{\mu^2 - \kappa^2}{\mu} \qquad (5)$$

Now, we have two different evanescent modes namely as $\psi_\mu$ and $\psi_{sp}$ in MWIFD. In early history of microwave ferrite studies, two different parallel travelling waves coupling phenomena were very well studied by H. Suhl [15] and J. R. Eshbach [16] for parametric ferrite amplifier purposes although the limitations of spin instabilities near the ferromagnetic resonances kept the problem challenging. Also, the parallel travelled waves couplings in terms of evanescent waves case was studied by S. E. Miller [17]. Since we have two different parallel evanescent waves in here, the successful propagating wave from evanescent waves coupling that may be expressed as follows, *the displacement of the surface wave of spin precession around the surface plasmon state that causes fluctuations of surface plasmon states and the vice versa*. The mathematical expression kept using Feynman's notation [12] for above statement and it is written as follows.

$$\frac{\partial \psi_\mu}{\partial z} = \beta \psi_{sp} = jk_\mu \psi_\mu \qquad (6)$$

$$\frac{\partial \psi_{sp}}{\partial z} = \beta \psi_\mu = jk_{sp0} \psi_{sp} \qquad (7)$$

Using equations (6) and (7), we obtain desired propagation constant for restored wave propagation by coupling evanescent waves inside waveguide, is shown below,

$$\beta^2 = -k_\mu k_{sp0} \text{ and } \beta = j\frac{\omega}{c}\sqrt{\varepsilon_{sp}\mu_f} = j\frac{\omega}{c}n_{sp} \qquad (8)$$

Equation (8), stands for a propagation constant of restored wave through MWIFD composite inside waveguide (see Fig.1), defined as $\beta = jk_{sp}$. Intuitively, now, the power flow along z-axis inside waveguide, which can be expressed as $S = \omega^2 c^{-2} A_m A_d$. Because magnetic vector potential $A_m$ of RF and introduced dielectric vector potential $A_d$ of spin flow precession are orthogonal to each other in nature. It implies that cross product of dielectric and magnetic vector potentials are also responsible to transfer power flow by coupled evanescent surface waves in waveguide.

Above analysis shows that one can change systems permittivity by using external magnetic field's influence while system permeability due to change. A pair nickel ferrite disks with 0.5 mm thickness was used to cover metal wire of aluminum foil. The table 1 represents the data summary for nickel ferrites used for MWIFD composite inside rectangular waveguide.

Table 1

| a | D | b | $4\pi M_s$ | $H_0$ | $\Delta f_{ferrite}$ | $\Delta f_{metal}$ |
|---|---|---|---|---|---|---|
| 2.3cm | 1.5mm | 2.2mm | 4kOe | 4.5kOe | 0.8GHz | 0.01GHz |
| 2.3cm | 1.5mm | 2.2mm | 4kOe | 5.5kOe | 0.8GHz | 0.01GHz |

The magnitudes of transmission coefficients $S_{21}$ measured using vector network analyzer is presented in Fig.3 and Fig. 4. According to various textbooks convention and technical definition, the $S_{21}$ is defined as $S_{21} = \frac{V(2)^+}{V_I(1)^+}$, where $V(2)^+$ = voltage measured at terminal (2) and $V_I(1)^+$ = voltage incident upon terminal (1). The $S_{21}$ is usually called "forward voltage gain" or "transmission coefficient", it is a transmission from right to left through the network.

In MWFID composite, the experimental data support the findings in equation (1) and (5). One cannot imagine the data in Fig.3 without accepting the concept the permittivity change under external magnetic fields influence due to spin flow of excited ferrite that sweeps the metal wires.

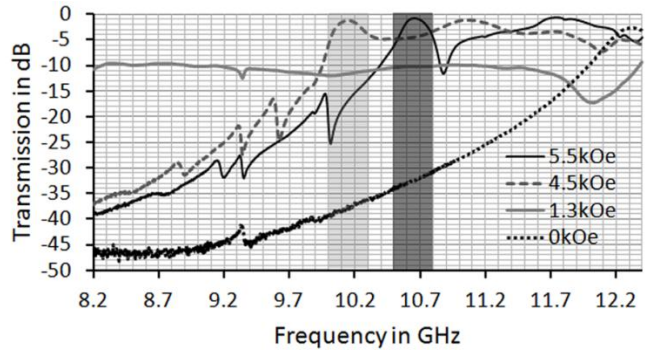

Fig. 3 Measured $S_{21}$ parameters showed correspondence to the positive and negative permeability and permittivity spectra in Fig.3 for 4.5kOe and 5.5kOe. The eliminated wave at 0kOe, and the eliminated wave was almost recovered by means of full transmission at highlighted bands in Fig.3. It implies that the negative permittivity in 0kOe becomes positive when the ferrite sees the external fields of 4.5kOe and 5.5kOe. Additionally, we measured a point at 1.3kOe field; it is consisted to the measurements in [7]. Here, nickel ferrite is used in MWIFD composite.

III. DISCRETE RADIATION BANDS AND DISPERSIVE CASMIR PRESURE FLIPS

The experimental measurement in Fig.3 confirmed the relationship between permeability and permittivity in MWIFD. Our analysis also indicates that multiple resonant sates existence in MWIFD. A pair YiG ferrite disks with a thickness ten mil each was purchased from Countis Laboratory to cover metal wires of aluminum foil. The table 2 represents data summary for MWIFD composite inside rectangular waveguide, where loading factor was into account to saturated magnetization due to very thin YIG ferrite disks were used in MWIFD. Interestingly, the experimentation in Fig.4 showed the double resonant states which were expected by our analysis.

Table 2

| a | d | b | $4\pi M_s$ | $H_0$ | $\Delta f_{ferrite}$ | $\Delta f_{metal}$ |
|---|---|---|---|---|---|---|
| 2.3cm | 2mm | 2.2mm | 1.2kOe | 2.6kOe | 0.03GHz | 0.01GHz |
| 2.3cm | 2mm | 2.2mm | 1.2kOe | 2.98kOe | 0.03GHz | 0.01GHz |

The measured magnitudes of transmission coefficient $S_{21}$ due to resonant permeability and permittivity spectra of MWIFD inside the waveguide are presented in Fig. 4, respectively. However, interpreting the Fig.4 is still difficult within scope of using conventional knowledge of metamaterials and magnetics alone. It is that has been difficult for us to interpret multiple pass bands by using published information thus far.

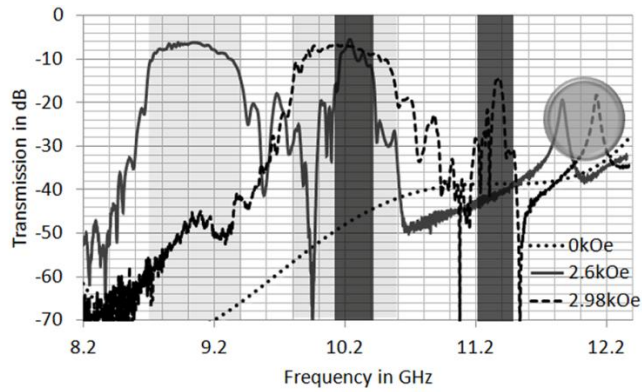

Fig. 4 Measured $S_{21}$ parameters showed the recreated pass band for positive refractive index and negative refractive index and they fairly linear dependent on external magnetic fields of 2.6kOe and 2.98kOe. The eliminated wave was recorded at 0kOe. The measurement shows that wider pass band may correspond to the positive refractive index while the narrower pass band corresponds to the negative refractive index. The grey circled region we don't know them yet. Here, YiG (yttrium iron garnet) is used in MWIFD. We recall this figure in section 4, where Josephson frequencies are used to interpret the discrete radiations in here.

The Casimir effect was introduced by H.B. G. Casimir and E. M. Lifshitz almost 60 years ago [18, 19]. The use of surface plasmon modes for Casimir forces was described recently by Munday et al. and Zao et al. [20, 21]. The effects of macroscopic quantum electrodynamics (QED) were also shown by Buhmann et al. [22] to a magneto-electric medium, where a dispersive Casimir force was shown in terms of permeability and permittivity of electromagnetic wave medium. Recently, a dynamic Casimir effect by using Josephson metamaterial was also reported in [23] by L. Lahteenmaki et al., and the report showed that energy correlated photon generation by touching the ground state energy in vacuum. All those findings suggest that there is a repulsive Casimir pressure effect in magneto-electric and metamaterial composites. For a composite of quasi 1D metal wires in ferrite disks described here; multiple reflections [7] for Figs.1 and 2 were considered using equation [9] below.

$$\Gamma(@\,0) = r(1 - re^{-jk_{sp}d} + r^2 e^{-j2k_{sp}d} - \cdots) \quad (9)$$

In equation (9), reflection configuration includes oscillating surface plasmon quanta in the first reflection of MWIFD. It resembles the characteristics of partition function of discrete microstates in statistical dynamics. For such matter of periodic metallic mirrors in metamaterialized ferrite composite structure in Fig.1 and 2, we share the idea of Lifshitz who extended Casimir force into medium [19], and the Casimir energy per unit area is designed in respect to Lifshitz and Milloni [19, 24] expressions that may now be written as,

$$E_S = -\frac{\hbar}{2} \int_0^{j\infty} d\zeta \int_{k=k_{sp}}^{\infty} \frac{kdk}{(2\pi)^2} \ln\left(\frac{r}{1+re^{-jk_{sp}d}}\right) \quad (10)$$

Where $\zeta = -j\omega$, and the k space has oscillating surface plasmon quanta that is represented by $k_{sp}$. By taking a condensed variable $\xi = j\frac{\omega}{c}n_{sp}d$, where $n_{sp}$ the refractive index of propagating surface plasmon in MWIFD and d is the width of each periodic metal wires. We have obtained a Casimir force for a periodic metal mirrors that lies between ferrite insulators as below.

$$\tilde{F}_S = \frac{\hbar c}{16\pi^2 d^4} \int_0^\infty \frac{1}{n_{sp}} \frac{\xi^3 r}{e^\xi + r} d\xi \qquad (11)$$

To test reasonableness of equation (11), let us first take a look for two extreme ideal cases of the problem (lossless cases, for example). If reflectivity to be r = -1 ($\varepsilon_{sp} \to \infty$) we have obtained exact Casimir force [18] that is for frequency dependent dielectric mirror ($\varepsilon_{sp} \to \infty$), shown in Table 3. If reflectivity to be r = 1 ($\mu_f \to \infty$) we have obtained exact Casimir force by Boyer [25] and Hushwater [26] that is for frequency dependent magnetic mirror ($\mu_f \to \infty$), also shown Table 3.

Table 3 Casimir forces for ideal medium and ideal cases

|  | r = -1 | r = 1 |
|---|---|---|
| Casimir forces (ideal medium) | $F_S = -\frac{\pi^2 \hbar c}{240 n_{sp} d^4}$ | $F_S = \left(\frac{7}{8}\right)\frac{\pi^2 \hbar c}{240 n_{sp} d^4}$ |
| Casimir forces (if ideal medium becomes air) | $F_S = -\frac{\pi^2 \hbar c}{240 d^4}$ | $F_S = \left(\frac{7}{8}\right)\frac{\pi^2 \hbar c}{240 d^4}$ |

For both cases, if refractive index of medium hovering close to zero ($n_{sp} \to \pm 0$), then positive or negative Casimir force becomes enormous Casimir pressures, which may be observed through the macroscopic systems. In order to work out equation (11) for a dissipative system, one requires refractive index of such system to remain a measurable quantity throughout due to Kramers-Kronig dispersion relationships [27]. This requirement prevents unnecessary gains by such systems in the calculation process. We use the following transformation in here $e^\xi = z$, which allows us to move the equation (11) into a complex domain, shown in equation (12) below.

$$\tilde{F}_S = \frac{\hbar c}{16\pi^2 d^4} \oint_c \frac{r}{n_{sp}} \frac{(\ln z)^3}{z+r} \frac{1}{z} dz \qquad (12)$$

$$\tilde{F}_s = \frac{\hbar c}{16\pi^2 d^4} \oint_c \frac{1}{n_{sp}} \left(\frac{1}{z} - \frac{1}{z+r}\right)(\ln z)^3 dz$$
$$= \tilde{F}_{S1} + \tilde{F}_{S2}$$

Where

$$\tilde{F}_{S1} = \frac{\hbar c}{8\pi d^4}\left[j\frac{(\ln z)^3}{n_{sp}}\right]_{z=0} \to \infty \quad \tilde{F}_{S2} = \frac{\hbar c}{8\pi d^4}\left[\frac{1}{j}\frac{(\ln z)^3}{n_{sp}}\right]_{z=-r}$$

This is the Casimir pressure effect from surface plasmon quanta in MWIFD described here, which is able to interpret obtained experimental data in Fig. 4 in terms of discrete radiation patterns. One easily finds discrete Casimir pressures flips by using equation (12), where the strict Cauchy integration technique was used for. The Casimir pressure flips may indicate instantaneous power release or absorption between quantized energy state transitions of surface plasmon quanta in MWIFD. By using usual convention of resonant states, it usually absorbs $S_{21}$ transmissions. Pure negative and positive refractive index is achieved below and beyond the resonant states which have the limits to interpret discrete pass bands in experimental observations in this paper. We do expect the repulsive Casmir pressure effect be further confirmed by showing such radiation in other experimental determination (other than power ratio measurement, $S_{21}$). Such radiation sources may be applied as principal radiation source in biological tissue detection techniques as well as renewable energy techniques.

### IV. JOSEPHSON FREQUENCIES AND CURRENTS

According to the recent reports in [9, 10, 11, 28, 29, 30 and 31], the ferromagnetic resonances and spin waves in ferrites they offer a new opportunity for creating electric flux or spin electromotive force by uniform spin excitations. In previous sections, this study presented induced electric flux associated surface wave propagation and the dispersive Casimir forces for the creation of potential electromagnetic waves radiations. Both Casimir force analysis and measured discrete radiation spectra in fig.4 show that MWIDS system develops quantized states. Although the metal wires in ferrite disks is considered a plasmon matter in terms of negative permittivity medium for the surface wave propagation, the metal wires is also the conductors who sees the induced electric flux by ferrite excitation. As result, we used the Schrödinger representation, where dynamics specified time dependent wave function and it is a convenient approach to deal with interaction between electrons in metal wires and electric flux induced by ferromagnetic resonance in ferrites. The Schrödinger equation is shown here as $i\hbar\partial_t|\psi(t)\rangle = \left(\hat{h}_0 + v_d(t)\right)|\psi(t)\rangle$, where $\hat{h}_0$ stands Hamiltonian operator for linear stationary electron in MWIFD, and $v_d(t) = \left(e_q v_d^0\right)e^{i\omega t}$ stands electron ($e_q$, electron charge) volt energy of an electron in metal wire via voltage induced by ferromagnetic resonance in MWIFD. This electron volt energy via voltage induced by ferromagnetic resonance in MWIFD must be as $h_0 \gg (e_q v_d^0)$, so it can be considered a time dependent perturbation in the Schrödinger equation. By using Fermi's Golden Rule [32], one obtains resonance excitation frequencies (radiation frequencies) as follows $\omega_{nj} = \frac{e_q}{\hbar}v_d^0$. This simply represents Josephson frequencies in the system which means electrons in the system (MWIFD) develop quantized discrete states by electric flux induced by ferromagnetic resonance in ferrite. According to the data in Fig. 4, we are able to calculate the voltages induced for resonant excitations as follows, 5.7µV at 8.8GHz, 6.7µV at 10.2GHz and 6.8µV at 10.4GHz. The order of magnitude of obtained data very well agrees with other reports in [10, 11, and 30]. According to the Fermi's Golden Rule, we have the transition probability as follows, $w_{j \to n} = \frac{2\pi}{\hbar} |\langle \psi_n | e_q v_d^0 | \psi_j \rangle|^2 \rho_{sp}(\omega_{nj})$, we will use it when we come to think system's radiation intensity in terms of energy matter. Over there $\rho(\omega_{nj})$, namely density of states. Let us make a reasonable estimate to the system's power creation for a possible transition by $\langle \psi_{200} | e_q e_d^0 \hat{z} | \psi_{210} \rangle = 3 a_{b0} e_q e_d^0$, where $a_{b0}$, Bohr radius. Suppose we have electric flux induced by ferromagnetic resonance as $e_d^0 = 5 \times 10^{-4} Vm^{-1}$ and density of states reached as $\rho_{sp}(\omega_{nj}) \propto 10^{11}$ (i.e., we know the density of free electron gas in a good conductor as $N \approx 8.48 \times 10^{22} cm^{-3}$ and one finds electron density in an excited surface plasmon to be as $\rho_{sp} \approx \sqrt{N} \propto 10^{11}$), then the peak voltage inside waveguide about to be 5Volt, which means about a 20mW microwave power flow inside waveguide of X-band possible. Apparently, it will be an enormous power creation if we achieve it. In near future, we shall discuss density of states for a study of radiation intensity matter in a separate paper. More recently, Barnes and Meakawa in their report[28] they showed the Berry phase inclusion in the Schrödinger equation due to spin electromotive force by ferrite and Meakewa also reported a Josephson current induced by ferromagnetic resonance in YIG [29]. An ISHE (inverse spin hall effect) report [30] the authors showed pulsed microwave excitation that creates greater voltage signature compared to the temporal evolution of directly excited spin wave mode in YIG. We also noted that the inverse spin hall by using spin torque dynamics was also in great interest to amplify microwave signals [31].

According to those experimental evidences and findings, we may be able to write the Schrödinger representation here in a new form as follows, $i\hbar \partial_{t'} |\psi(t')\rangle e^{i\frac{e_q}{\hbar} \int_0^{t'} v_d^0(t) dt} = \hat{h}_0 |\psi(t')\rangle e^{i\frac{e_q}{\hbar} \int_0^{t'} v_d^0(t) dt}$, where $e_q$ electron charge. One assumes we should be able to reach a phase condition where the phase component in above Schrödinger representation has to be integer value as $\frac{e_q}{\hbar} \int_0^{t'} v_d^0(t) dt = 2\pi n$, where n, integer number and Planck's constant as $\hbar = 6.58 \times 10^{-16} eV \cdot s$. If so, one obtains the quantized current flow in metal wire as follows, $i_{ld} = -n \frac{e_q}{L} \left( \frac{h}{e_q^2} \right) = -n \frac{e_q}{L} R_{k\_qh} = -n e_q \left( \frac{R_{k\_qh}}{L} \right)$, where $R_{k\_qh}$, Klitzing resistance and L, inductance. In our system of MWIFD the charge in metal wires has to be conserved, so one obtains induced charge distribution in metal wires as $q = q_{ld} e^{-\frac{R_{qh\_k}}{L} t'} = (-n e_q) e^{-\frac{R_{qh\_k}}{L} t'}$, where $q_{ld} = -n e_q$. The above formulations present us two interesting phenomena. One of them is that one can achieve quantized current flow at room temperature, which may be understood as Josephson current [29] at room temperature. For example, in fig. 4, unlike those resonance radiation spectra, they are also many unusual deep absorption cuts in $S_{21}$, one can see those deep absorption cuts at 9.9GHz, 11.1GHz and 11.3GHz in fig.4. They may be the frequencies spectra where we reached the conditions for Josephson currents. Perhaps, they are the frequencies where DC Josephson currents available. Furthermore, our charge distribution shows that induced charge on metal wires decays very fast if it is not located on negative inductance spectrum, which may be an additional indication why pulsed microwave excitation better for induced voltage creation [30].

## V. CONCLUSION

Ferrite materials researches scientists reported inverse spin hall effect, spin electromotive force, electric flux induced by ferromagnetic resonances (FMR) and every other voltage induced by FMR. They all clearly related to the electromotive force, which can be represented by E field. We found electric flux induced by ferromagnetic resonance in a ferrite which very much finds agreement with other reports. In this paper, the concept of a dielectric vector potential was also introduced into the system of MWIFD, to generate a new set of Maxwell equations to recover evanescent surface plasmon due to flow by bulk or surface spin waves in ferrite. The dispersive Casimir pressure is derived by using surface plasmon quanta which removes atypical phenomenon of negative absorption from a composite of 1D metal wires structure in ferrite disks. The composite showed both negative permittivity and permeability simultaneously to be a metamaterial. The Casimir pressure flips from attractive state to repulsive state is referred to the cause of radiation from surface plasmon quanta in this paper. Evidently, repulsive radiation pressure from Casimir pressure flips due to energy state transitions of surface plasmon quanta in MWIFD, which is understood as a cause mechanism to interpret radiation pattern of measurements in terms of classical interpretation. Moreover, Schrödinger representation was used here to derive Josephson frequency and Josephson current for the MWIFD system. The Josephson frequency allowed us to extrapolate the voltage induced by ferromagnetic resonance in YIG. We also realized that density of states is important to the radiation intensity in terms of energy matter. This study may be beginning of a new quest in searching high power radiation sources by using ferrite based metamaterials although quantization of the electrodynamics in metamaterial medium is another approach.

Author gratefully acknowledges that experimental measurements in this paper used Microwave Materials Laboratory facilities at Northeastern University, Boston MA.